\documentstyle[prd,aps]{revtex}
\def\slash {\not\!}
\def\ftriangle {\hskip -1.5pt}
\begin{document}

%\iffalse 
\draft
\title{Compositeness condition for dynamically induced gauge theories}
\author{Keiichi Akama}
\address{Department of Physics, Saitama Medical College,
                     Kawakado, Moroyama, Saitama, 350-04, Japan}
\author{Takashi Hattori }
\address{Department of Physics, Kanagawa Dental College,
                        Inaoka-cho, Yokosuka, Kanagawa, 238, Japan}
%\date{\today}
\maketitle
\begin{abstract}
We show that the compositeness condition 
	for the induced gauge boson in the four-fermion interaction theory
 	actually works beyond the one-loop approximation.
The next-to-leading contributions are calculated,
	and turn out to be  reasonably suppressed,
	so that the leading-order approximation is justified.
\end{abstract}
%\fi 

From theoretical points of view,
	the gauge fields can be dynamically induced in various systems
	even without preparing them 
	as fundamental objects \cite{QED} -- \cite{Kik}.
In fact we have many signatures for it.
In some models, gauge fields are induced as composites
	via nonlinear interactions of matters \cite{QED} -- \cite{Hid}.
They are applied to QED \cite{QED}, hadron physics \cite{Had},
	the electroweak theory \cite{EW,TCA,EW2}, 
	QCD \cite{TCA,QCD}, induced gravity \cite{IG,IG2} and 
	the theory of hidden local symmetry \cite{Hid}.
In quantum theory on constrained systems 
	embedded in higher dimensional ones, 
	gauge fields arise as 
	induced connections. 
They are realized for particles confined in a tube or a layer \cite{QM},
	and are applied to the embedding models of the 
	spacetime \cite{Emb},
	providing a unification scenario 
	of the gauge theory and gravitation.
Recently emergence of gauge structure attracted attention
	in connection with quantization 
	on non-trivial sub-manifolds \cite{QM2}.
Gauge fields are induced in more general systems
	associated with geometric (Berry) phases \cite{GP}.
The effects are observed or expected in
	the optical, molecular, solid state, nuclear, 
	elementary particle and cosmological systems \cite{GP2,GP3},
	and, in some models, they are taken as 
	the origin of the dynamical gauge fields \cite{Kik}.

In these models, we expect that 
	the gauge fields become dynamical propagating fields
	through the quantum fluctuations of 
	the matters \cite{QED} -- \cite{Hid}, \cite{Emb}, \cite{Kik}.
A typical model realizing the mechanism is
	the four-fermion interaction theory, 
	where a gauge field is induced as a composite 
	of a fermion-anti-fermion pair \cite{QED} -- \cite{EW2}.
The model is non-renormalizable in four dimensions, and we fix the momentum
	cutoff at some large but finite value. 
Then the strong coupling limit of the model is known to be equivalent to 
	the gauge theory (with the cutoff fixed) under the compositeness 
	condition, $Z_3=0$ \cite{LM,CC},
	where $Z_3$ is the wave function renormalization constant 
	of the gauge field.
Thus what is urgent for the induced gauge theory
	is to work out the compositeness condition.

Though there exist several formal arguments 
	on the compositeness conditions at higher order \cite{LM}, 
	so far no reliable investigations have been performed 
	beyond the one loop approximation.
In this paper, we show that, in spite of apparent problems on the way, 
	the compositeness condition in the abelian gauge theory 
	(with finite cutoff) actually works at the non-trivial higher order. 
As is seen below, 
	the compositeness condition itself spoils
	the ordinary perturbation expansion 
	in terms of the coupling constant.
A possible alternative is 
	the $1/N$-expansion
	with the number $N$ of the fermion species. 
The next-to-leading order already involves
	infinite series of diagrams,
	which give rise to a logarithmic singularity in the coupling constant.
It is remarkable, however, that 
	the compositeness condition is successfully solved 
	to give a surprisingly simple solution (Eq.\ (\ref{er1}) below).

In a separate paper \cite{AkCC}, a similar investigation recently performed 
	for the composite scalar fields. 
It was, however, not really successful because
	the next-to-leading-order corrections 
	are too large {\it e.g.} for $N=3$,
	and consequently the lowest order approximation was not acceptable. 
(See also the related arguments in \cite{ZJ} -- \cite{Pallante}.)
In contrast, the next-to-leading-order correction in the present model
	is reasonably suppressed by an extra factor of 
	$ \epsilon = (4-d)/2 $ 
	($d$ is the number of the space-time dimensions),
	and the lowest order approximation is justified.

We consider the strong coupling limit $f\rightarrow \infty $ of
	the vector-type four-fermion interaction theory 
	for the fermion field $\psi _j$ ($j=1,2,\cdots ,N$) \cite{QED}:
\begin{eqnarray} 
{\cal L}_{\rm 4fermi}
	=\sum _j{\overline \psi }_j \left(   i {\slash \partial }-m_j \right)   \psi _j
	 -f \left(  \sum _j {\overline \psi }_j \gamma _\mu Q_j \psi _j \right)  ^2,		\label{L4f}
\end{eqnarray} 
	where 
	$f$ is the coupling constant, 
	$m_j$ is the mass of $\psi _j$, and
	$Q_j$ is the constant which will be identified with
	the charge of $\psi _j$.
The system described by the Lagrangian ${\cal L}_{\rm 4fermi}$ 
	is equivalent to that with the Lagrangian 
\begin{eqnarray} 
{\cal L}'_{\rm 4fermi}
	=\sum _j{\overline \psi }_j 
	\left(   i {{\slash \partial }}-m_j- Q_j {\slash \ftriangle A} \right)   \psi _j
	 +{ 1 \over 4f } \left(   {A_\mu } \right)  ^2 .		\label{La0}
\end{eqnarray} 
with the auxiliary field ${A}_\mu $. 
We take the limit $f\rightarrow \infty $, and discard the last term hereafter. 
\begin{eqnarray} 
	{\cal L}'_{\rm 4fermi}
	=\sum _j\overline \psi _j \left( i\slash \partial -m_j- Q_j {\slash \ftriangle A} \right)   \psi _j.      \label{La}
\end{eqnarray} 
Then ${\cal L}'_{\rm 4fermi}$ becomes invariant under abelian gauge 
	transformations.
The kinetic term of $A_\mu $ is absent in (\ref{La}),
	and is induced through quantum loop diagrams.
In the previous works \cite{QED} -- \cite{IG2}, however, 
	only one-fermion-loop diagrams are examined.
Then arises the question how we can incorporate the multi-loop diagrams
	including induced boson lines.
The naive ways suffer from unmanageable divergences 
	due to the non-renormalizability of the original model.
If we first derive the effective action by performing $\psi $ integration, 
	it involves non-renormalizable higher power terms in $A_\mu $.
If we use the Schwinger-Dyson or the renormalization group equations,
	we can incorporate only a particular class of higher-order diagrams.
There is no guarantee that the discarded diagrams are really negligible.

In order to avoid these difficulties,
	we need to renormalize the various quantities in a systematic way.
For this purpose,
	we compare (\ref{La}) with the fully renormalized version
	of the elementary gauge field theory:
\begin{eqnarray} 
	{\cal L}_{\rm gauge}
	= \sum _j {\overline \psi }_{{\rm r}j} 
	\left(   i Z_{2j} {{\slash \partial }}-Z_{mj}m_{{\rm r}j} 
	-Z_{1j} e_{\rm r} Q_j{\slash \ftriangle A}_{\rm r}\right)   \psi _{{\rm r}j}
	 -{ 1 \over 4 } Z_3 \left(   F_{\rm r}^{\mu \nu } \right)  ^2 ,
	                                                        \label{Lgr}
\end{eqnarray} 
	where 
	$\psi _{{\rm r}j}$, $A_{\rm r}^\mu $, $e_{\rm r}$ and $m_{\rm r}$ 
	are the renormalized fermion field, gauge field,
	coupling constant and mass of $\psi _{{\rm r}j}$, respectively,
	and $F_{\rm r}^{\mu \nu }= 
	\partial ^\mu A_{\rm r}^\nu - \partial ^\nu A_{\rm r}^\mu $. 
In (\ref{Lgr}), $Z_{1j}$, $Z_{2j}$, $Z_3$ and $Z_{mj}$ 
	are the renormalization constants,
	which relate the renormalized quantities
	$\psi _{{\rm r}j}$, $A_{\rm r}^\mu $, $e_{\rm r}$ and $m_{\rm r}$ 
	to the corresponding bare quantities
	$\psi _{{\rm b}j}$, $A_{\rm b}^\mu $, $e_{\rm b}$ and $m_{\rm b}$ by
\begin{eqnarray} 
	A_{\rm b}^\mu = \sqrt {Z_3} A_{\rm r}^\mu , \ \ 
	\psi _{{\rm b}j} = \sqrt {Z_{2j}} \psi _{{\rm r}j} , \ \ 
	{Z_{2j}}m_{{\rm b}j}= {Z_{mj}} m_{{\rm r}j}, \ \ 
	Z_{2j} \sqrt {Z_3} e_{\rm b} = Z_{1j} e_{\rm r} . 	\label{Zdef}
\end{eqnarray} 
As is well known the Ward identity implies that $Z_{1j}= Z_{2j}$.
Now we find that the Lagrangian (\ref{Lgr}) of the 
	elementary gauge field theory coincides with the Lagrangian 
	(\ref{La}) of the four-fermion theory with $f\rightarrow \infty $, 
	if we 
impose the compositeness condition \cite{LM,CC}
\begin{eqnarray} 
	Z_3 = 0 ,						\label{CC}
\end{eqnarray} 
	and identify the fields and masses as 
	${A_\mu }= e_{\rm r} A_{{\rm r}\mu }$, 
	$\psi _j= \sqrt {Z_{2j}} \psi _{{\rm r}j}=  \psi _{{\rm b}j}$ and
	$m_j=({Z_{mj}}/Z_{2j})m_{{\rm r}j}=m_{{\rm b}j}$. 
The condition (\ref{CC}) imposes a relation 
	among the parameters in the elementary gauge field theory so that 
	it reduces to the four-fermion theory. 
The coupling constant $e_{\rm r}$ is related to the cutoff by this relation.
In terms of the bare quantities, 
	the condition (\ref{CC}) imply the singular limit 
	$A_{{\rm b}\mu }=0$ and $e_{\rm b}=\infty $. 
It causes, however, no harm to the renormalized quantities.

Thus, the strong coupling limit of
	the four-fermion theory is  equivalent to the elementary gauge
	theory (with the cutoff fixed) under the compositeness condition 
	(\ref{CC}), as far as we assume the same fixed cutoff 
	for both theories.
This is simply because everything is to be calculated based on the same 
	Lagrangian and the same cutoff methods.
It does not mean that 
	the four-fermion theory is equivalent to the renormalizable and 
	nontrivial gauge theory at the infinite cutoff limit.
In this limit, the former gives a trivial free theory with vanishing 
	coupling constant (see below).
To avoid the absurdity, we fix the cutoff at some large but finite value,
	which we take as a real physical one.
Then, they are equivalent in the sense stated above.
This equivalence implies that 
	we can in principle calculate the quantities
	in the four-fermion theory at any higher order
	by using the established methods in the elementary gauge theory.
Thus, what is urgent in the former is 
	to work out the compositeness condition $ Z_3=0 $ 
	and to solve it for $e_{\rm r}$.

This formal equivalence has been known for a long time \cite{LM},
	but it is not the whole story, 
	because the compositeness condition itself 
	spoils the usual perturbation expansion in $e_{\rm r}$. 
To see this, let us consider the lowest order approximation.
For a while we assume that the charge $Q_j$ are equal to each other for 
	simplicity. 
The one-loop diagram in Fig.1 gives 
\begin{eqnarray}  
	Z_3=1-{e_{\rm r}^2 N \over 12\pi ^2\epsilon },   \label{Z3}
\end{eqnarray}  
	where we have used the dimensional regularization. 
Then the compositeness condition $Z_3=0$ is solved to give
\begin{eqnarray} 
	e_{\rm r}^2 = { 12 \pi ^2 \epsilon \over N } . \label{er}
\end{eqnarray}  
This is the known result in the naive chain approximation \cite{QED}.
From (\ref{er}) we can see that the theory becomes trivial as 
	$ \epsilon \rightarrow 0 $. 
So we fix $\epsilon $ at some small but non-vanishing value. 
We take the dimensional regularization as an approximation to some realistic 
	momentum cutoff $\Lambda $ . 
It is well known that $\epsilon $ corresponds to $ 1/ {\rm ln } \Lambda ^2 $ 
	in real cutoff scheme at this order. 
Suppose we insert $l$ fermion loops to boson lines in a higher order diagram,
	it will acquire an extra factor of
	$O(({e_{\rm r}^2 N/12\pi ^2\epsilon })^l)$.
If we use the leading order relation (\ref{er}), the extra factor is $O(1)$,
	and the higher order is not suppressed.
Thus the perturbation expansion in $e_{\rm r}$ fails. 

Let us seek for alternative expansion parameters.
Suppose that a self-energy diagram involves $l_{\rm t}$ loops in total 
	and $l_{\rm f}$ fermion loops.  
Then a power counting shows that it behaves like
\begin{eqnarray}  
	\left(  {e_{\rm r}^2\over \epsilon }\right)  ^{l_{\rm t}}
	N^{l_f} 
	\sim { 1 \over N^{l_t - l_f} } , 
								\label{behav}
\end{eqnarray}  
where we have used the lowest order relation (\ref{er}), 
	and we can expand the quantity in $1/N$.
Even though the $Q_j^2$'s are not equal, 
	the hierarchy of the magnitude are maintained,
	as far as $N$ is large and $Q_j^2\approx O(1)$ for all $j$.
Thus we can expand the quantity according to this hierarchy.

The next-to-leading contributions in $1/N $-expansion of $Z_3$ 
	are given by the diagrams in Figs.\ 3(a) and (b).
The lowest-order self-energy part in Fig.\ 1 for a gauge boson with momentum 
	$q_\mu $ is given by 
\begin{eqnarray}  
	c_1{ e^2 N \over 12 \pi ^2 \epsilon }(-q^2)^{-\epsilon }
	( -g_{\mu \nu }q^2+q_\mu q_\nu )+O(m_j^2) ,
\end{eqnarray}  
	where
\begin{eqnarray}  
	c_1=6(2\pi )^\epsilon \Gamma (1+\epsilon )B(2-\epsilon , 2-\epsilon ) 
	\longrightarrow 1, 
\end{eqnarray}  
	as $ \epsilon \rightarrow 0 $. 
The $O(m_j^2)$ terms do not finally contribute to the divergence of 
	Fig.\ 3(a) and (b).
Then, the fermion self energy part in Fig.\ 2 with momentum $k_\mu $ becomes 
\begin{eqnarray}  
	c_2{ e^2 \over 16\pi ^2 \epsilon }
	\left( { -e^2 N \over 12\pi ^2 \epsilon } \right) 
	{(-k^2)^{-\epsilon (l+1)} \over (l+1)}{\slash k}+O(m_j),
	                                                    \label{selff}
\end{eqnarray}  
	where $l$ is the number of the lowest self-energy parts inserted 
	on the internal boson line, and 
\begin{eqnarray}  
	c_2 = 2(4\pi )^\epsilon (1-\epsilon )
	{ \Gamma (1+\epsilon (l+1))B(2-\epsilon , 1-\epsilon (l+1)) \over 
	         \Gamma (1+\epsilon l) } c_1^l \longrightarrow 1,
\end{eqnarray}  
	as $\epsilon \rightarrow 0 $. 
Using (\ref{selff}) we obtain the following expression for the boson-self 
	energy part in Fig.\ 3 (a) with momentum $p_\mu $:
\begin{eqnarray}  
	\Pi _{\mu \nu }^{({\rm a})}= 
	c_3{ e^2 N \over 24\pi ^2 \epsilon } { e^2 \over 16\pi ^2 \epsilon }
	\left( { -e^2 N \over 12\pi ^2 \epsilon } \right) ^l 
	{ (-p^2)^{-\epsilon (l+2)} \over (l+1)(l+2) }
	\left( { 2-\epsilon (l+3) \over 1-\epsilon (l+2) }
	      g_{\mu \nu }p^2 - 2p_\mu p_\nu \right) ,
	                                                   \label{Pia}
\end{eqnarray}  
	up to $O(m_j^2)$ terms, where
\begin{eqnarray}  
	c_3= 6(2\pi )^\epsilon 
	     { \Gamma (1+\epsilon (l+2)) \over \Gamma (1+\epsilon (l+1)) }
	     B( 2-\epsilon , 2-\epsilon (l+2))c_2
	\longrightarrow 1
\end{eqnarray}  
	as $\epsilon \rightarrow 0$. 
To calculate the self-energy part in Fig.\ 3(b), we rewrite the part 
\begin{eqnarray}  
	{1 \over {\slash k}+{\slash p}}\gamma _\rho 
	{1 \over {\slash q}+{\slash p}},
\end{eqnarray}  
	in the integrand into the sum of the four parts
\begin{eqnarray}  
	& & D_1= {1 \over {\slash k}+{\slash p}}\gamma _\rho {\slash q},
	\ \ \ \ \ \ \ \ \ \ \ \ \ \ \ \ \ \ 
	    D_2= {1 \over {\slash k}}\gamma _\rho {1 \over {\slash q}}
	     {\slash p} {1 \over {\slash q}+{\slash p}},             \cr 
	& & D_3= {1 \over {\slash k}+{\slash p}} {\slash p}
	         {1 \over {\slash k}} \gamma _\rho 
	         {1 \over {\slash q}} {\slash p} {1 \over {\slash q}},
	\ \ \ \ \ \ \ \ \ \ 
	    D_4= {1 \over {\slash k}+{\slash p}} {\slash p}
	         {1 \over {\slash k}} \gamma _\rho 
	         {1 \over {\slash q}} {\slash p} {1 \over {\slash q}}
	         {\slash p} {1 \over {\slash q}+{\slash p}},
\end{eqnarray}  
	and define the corresponding self-energy parts by 
	$\Pi _{\mu \nu }^{({\rm b}1)}$, $\Pi _{\mu \nu }^{({\rm b}2)}$, 
	$\Pi _{\mu \nu }^{({\rm b}3)}$ and $\Pi _{\mu \nu }^{({\rm b}4)}$, 
	respectively. 
Then, we obtain
\begin{eqnarray}  
	\Pi _{\mu \nu }^{({\rm b}1)} = \Pi _{\mu \nu }^{({\rm b}2)}
	= & & c_3{ e^2 N \over 24\pi ^2 \epsilon }
	        { e^2 \over 16\pi ^2 \epsilon }
	 \left( {-e^2 N \over 12\pi ^2 \epsilon } \right) ^l
	 { (-p^2)^{-\epsilon (l+2)} \over (l+1)(l+2) }           \cr 
	& & \times 
	\left( {1-\epsilon \over 1-\epsilon (l+2)} g_{\mu \nu }p^2 - 
	      \left(  1-\epsilon (l+1) \right) p_\mu p_\nu \right) ,
	                                                       \label{Pib12}
\end{eqnarray}  
\begin{eqnarray}  
	\Pi _{\mu \nu }^{({\rm b}3)}=
	{ e^2 N \over 24\pi ^2 \epsilon }{ e^2 \over 16\pi ^2 \epsilon }
	\left( {-e^2 N \over 12\pi ^2 \epsilon } \right) ^l
	{ (-p^2)^{-\epsilon (l+2)} \over (l+2) }
	\left( 4 \epsilon g_{\mu \nu }p^2 
	              - 10 \epsilon p_\mu p_\nu +O(\epsilon ^2) \right) ,
	                                                       \label{Pib3}
\end{eqnarray} 
\begin{eqnarray} 
	\Pi _{\mu \nu }^{({\rm b}4)}=O(1/\epsilon ^l) 			\label{Pib4}
\end{eqnarray}  
	 up to $O(m_j^2)$.
From (\ref{Pia}) and (\ref{Pib12}) -- (\ref{Pib4})
	we can see that the leading divergent 
	parts of $\Pi _{\mu \nu }^{({\rm a})}$, 
	$\Pi _{\mu \nu }^{({\rm b}1)}$ and $\Pi _{\mu \nu }^{({\rm b}2)}$ 
	cancel out and in total, the next-to-leading divergent parts 
	contribute the term 
\begin{eqnarray}  
	-{e_{\rm r}^4 N \over 64\pi ^4 \epsilon (l+2)}
	\left(  -{e_{\rm r}^2 N \over 12\pi ^2\epsilon }\right)  ^l
                                                            \label{higher}
\end{eqnarray}  
	to $Z_3$. 
Taking into account the counter terms for sub-diagram divergences, we get
\begin{eqnarray}  
	-\sum _{k=0}^l{\pmatrix{l\cr k}}
	{(-1)^{l-k}e_{\rm r}^4 N \over 64\pi ^4 \epsilon (k+2)}
	\left(  -{e_{\rm r}^2 N \over 12\pi ^2\epsilon }\right)  ^l 
	=-{e_{\rm r}^4 N \over 64\pi ^4 \epsilon (l+1)(l+2)}
	\left(  {e_{\rm r}^2 N \over 12\pi ^2\epsilon }\right)  ^l.
\end{eqnarray}  
Note that only the counter terms 
	for the gauge boson self-energy parts contribute, 
	while contributions from those 
	for the fermion self-energy parts in Figs.\ 3(a) 
	and the vertex parts in Figs.\ 3(b) 
	cancel each other due to the Ward identity.
Performing the summation over $l$, 
	and adding it to the lowest order contribution (\ref{Z3}),
	we finally get
\begin{eqnarray}  
	Z_3=1-{e_{\rm r}^2 N \over 12\pi ^2 \epsilon  } 	
	-{3e_{\rm r}^2 \over 16\pi ^2 }
	\left[  1+\left(  1-{12\pi ^2\epsilon \over e_{\rm r}^2 N }\right)  
	\ln\left(  1-{e_{\rm r}^2 N \over 12\pi ^2\epsilon }\right) \right]  .              	\label{Z3n}
\end{eqnarray}  

Now the compositeness condition at this order
	is given by $Z_3=0$ with the expression in (\ref{Z3n}).
Though it involves a logarithmic singularity, 
we can solve it by iteration to get the surprisingly simple solution 
\begin{eqnarray}  
	e_{\rm r}^2 = { 12 \pi ^2 \epsilon \over N }\left[  1-{9\epsilon \over 4 N }\right]  .	\label{er1}
\end{eqnarray}  
Note that the correction term 
	(the second term in the square bracket in (\ref{er1}))
	is suppressed by the extra factor $\epsilon $, 
	which justify the lowest order approximation in the present model.
The origin of the suppression factor is traced back 
	to the gauge cancellation of the leading divergences
	of Figs.\ 3(a) and (b).
It is easy to extend the argument to the case where the fermions 
	$\psi _j $ have different electric charges $Q_j$. 
The result is 
\begin{eqnarray}  
	e_{\rm r}^2 = { 12 \pi ^2 \epsilon \over \sum _j Q_j^2 }
	\left[  1-{9\epsilon \sum _j Q_j^4\over 4(\sum _j Q_j^2)^2}\right]  .
	                                                        \label{ern}
\end{eqnarray}  
A simple minded application of (\ref{ern}) to QED with three generations
	indicates that $\epsilon \approx 6\times 10^{-3}$, 
	and the next-to-leading-order correction amounts to only 0.1\%.

This is in contrast with the large correction terms 
	for the coupling constants $g_{\rm r}$ and $\lambda _{\rm r}$
	of the composite scalars \cite{AkCC}:
\begin{eqnarray}  
	g_{\rm r}^2={16\pi ^2\epsilon \over N_{\rm c}}
	\left[  1-{1\over N_{\rm c}}\right] ,\ \ \ \ 
	\lambda _{\rm r}={16\pi ^2\epsilon \over N_{\rm c}}
	\left[  1-{10\over N_{\rm c}}\right] .\label{sol}
\end{eqnarray}  
where $N_{\rm c}$ is the number of the colors.
The correction terms in (\ref{sol}) are too large
	for {\it e.\ g.} $N_{\rm c}=3$, 
	which deflates the validity of the lowest order approximation itself.

Now we discuss extensions to non-abelian gauge theories.
Consider $SU(N_{\rm c})$ gauge theory with $N_{\rm f}$ ungauged flavors.
The total number of the fermion species is $N=N_{\rm c}N_{\rm f}$.
If all the degrees of freedom are gauged, {\it i.\ e.} if $N_{\rm f}=1$,
	the gauge boson self-energy diagram with a gauge boson loop 
	behaves like $g^2N^2/\epsilon \sim N$.
Those with the more gauge boson loops increase the more severely with $N$.
Thus the scenario to expand $Z_3$ in terms of $1/N$ fails.
On the other hand, if $N_{\rm c}$ is small and $N_{\rm f}$ is large,
	the gauge boson loops appear at one order behind
	and we can apply the above method in an almost parallel way.
This may be true for the case 
	of the weak bosons, where 
	$N_{\rm c}=2$ and $N_{\rm f}=12$,
	while we can not apply it to the QCD since 
	$N_{\rm f}=6$ is not sufficiently large compared with $N_{\rm c}=3$. 

In conclusion, we have demonstrated 
	the induced gauge fields in four-fermion theories  
	beyond the one-loop approximation
	by actually deriving the next-to-leading-order correction 
	to the coupling constants via the compositeness condition. 
The results are based on a definite expansion
 	in spite of non-renormalizability of the original model 
	and, in principle, we can calculate further higher order corrections.  
The correction term is reasonably suppressed
	so that the leading approximation is justified.
The results should seriously be taken into account 
	in the models of hadrons, weak bosons, and so forth.
We expect that it can also be applied 
	to the theories of induced gravity, hidden local symmetry,
	induced connections and geometric phases. 
For the models of induced connections or geometric phases,
	the present method amounts to quantizing the wave function of  
	the slow variables (see \cite{GP}) also.
The effective Lagrangian for them is essentially given by 
	(\ref{La}) where the coordinate $x$ corresponds to the slow 
	variable, $\psi _j$, to the wave function, and $A_\mu $, as the 
	induced connection. 
Then the induced gauge fields 
	become dynamical through the quantum effects of $\psi _j$, 
	just in the same way as described above. 
We wish that further studies will confirm the theoretical foundations 
	of the ideas
	and their effects will be observed in the future experiments.

We would like to thank Professors H. Terazawa, T. Matsuki, 
	I. Tsutsui and H. Mukaida 
	for invaluable discussions. 

%\newpage
\vskip 10mm
\noindent 
FIGURE CAPTIONS 

\noindent 
Fig.\ 1 \ \ The self-energy diagram which contribute to $Z_3$
	at the lowest order. 
The solid and wavy lines indicate the fermion and gauge boson propagators, 
	respectively. 

\noindent 
Fig.\ 2 \ \ The fermion self-energy diagram at the next-to-leading order 
	in $1/N$.
The solid and wavy lines indicate the fermion and gauge boson propagators, 
	respectively. 

\noindent 
Fig.\ 3 \ \ The gauge boson self-energy diagrams which contribute to $Z_3$
	at the next-to-leading order in $1/N $.
The solid and wavy lines indicate the fermion and gauge boson propagators, 
	respectively.

\end{document}